\documentclass[onecolumn,journal]{IEEEtran}
\usepackage{amsfonts}
\usepackage{amsmath}
\usepackage{amsthm}
\usepackage{amssymb}
\usepackage{graphicx}
\usepackage[T1]{fontenc}
\usepackage{supertabular}
\usepackage{longtable}
\usepackage[usenames,dvipsnames]{color}
\usepackage{bbm}
\usepackage{fancyhdr}
\usepackage{breqn}
\usepackage{fixltx2e}
\usepackage{capt-of}
\usepackage{tikz}
\usetikzlibrary{matrix}
\usepackage{endnotes}
\usepackage{soul}
\usepackage{marginnote}
\usepackage{authblk}

\newcommand{\mathsym}[1]{}
\newcommand{\unicode}[1]{}

\usepackage{pdfpages}
\usepackage{enumitem}
\usepackage{multicol}

\usepackage{float}

 \usepackage{xcolor}
 
\usepackage[framemethod=TikZ]{mdframed}
\usepackage[framemethod=TikZ]{mdframed}
\usepackage{framed}

\mdfsetup{%
	outerlinewidth=1,skipabove=20pt,backgroundcolor=yellow!50, outerlinecolor=black,innertopmargin=0pt,splittopskip=\topskip,skipbelow=\baselineskip, skipabove=\baselineskip,ntheorem,roundcorner=5pt}

\mdtheorem[nobreak=true,outerlinewidth=1,
backgroundcolor=yellow!50, outerlinecolor=black,innertopmargin=0pt,splittopskip=\topskip,skipbelow=\baselineskip, skipabove=\baselineskip,ntheorem,roundcorner=5pt,font=\itshape]{result}{Result}

\mdtheorem[nobreak=true,outerlinewidth=1,
backgroundcolor=yellow!50, outerlinecolor=black,innertopmargin=0pt,splittopskip=\topskip,skipbelow=\baselineskip, skipabove=\baselineskip,ntheorem,roundcorner=5pt,font=\itshape]{theorem}{Theorem}

\mdtheorem[nobreak=true,outerlinewidth=1,
backgroundcolor=gray!10, outerlinecolor=black,innertopmargin=0pt,splittopskip=\topskip,skipbelow=\baselineskip, skipabove=\baselineskip,ntheorem,roundcorner=5pt,font=\itshape]{remark}{Remark}

\mdtheorem[nobreak=true,outerlinewidth=1,
backgroundcolor=pink!30, outerlinecolor=black,innertopmargin=0pt,splittopskip=\topskip,skipbelow=\baselineskip, skipabove=\baselineskip,ntheorem,roundcorner=5pt,font=\itshape]{quaestio}{Quaestio}

\mdtheorem[nobreak=true,outerlinewidth=1,
backgroundcolor=yellow!50, outerlinecolor=black,innertopmargin=5pt,splittopskip=\topskip,skipbelow=\baselineskip, skipabove=\baselineskip,ntheorem,roundcorner=5pt,font=\itshape]{background}{Background}

\usepackage{hyperref}

\begin{document}
\title{\color{Brown}{Comment on: A systematic review and meta-analysis of published research data on COVID-19 infection-fatality rates} \\
\vspace{-0.35ex}}
\author{Chen Shen$^1$, Derrick Van Gennep$^1$, Alexander F. Siegenfeld$^{1,2}$, and Yaneer Bar-Yam$^1$\\ 
$^1$New England Complex Systems Institute, Cambridge, MA, USA\\
$^2$Department of Physics, Massachusetts Institute of Technology, Cambridge, MA, USA \\
November 26, 2020\\
  \vspace{-14ex}  

\bigskip
\bigskip
\textbf{
}
 }
    
\maketitle
\flushbottom 


\thispagestyle{empty} 




\renewcommand{\thefootnote}{\fnsymbol{footnote}}

{\bf
The infection fatality rate (IFR) of COVID-19 is one of the measures of disease impact that can be of importance for policy making. Here we show that many of the studies on which these estimates are based are scientifically flawed for reasons which include: nonsensical equations, unjustified assumptions, small sample sizes, non-representative sampling (systematic biases), incorrect definitions of symptomatic and asymptomatic cases (identified and unidentified cases), typically assuming that cases which are asymptomatic at the time of testing are the same as completely asymptomatic (never symptomatic) cases. Moreover, a widely cited meta-analysis misrepresents some of the IFR values in the original studies, and makes inappropriate duplicate use of studies, or the information from studies, so that the results that are averaged are not independent from each other. The lack of validity of these research papers is of particular importance in view of their influence on policies that affect lives and well-being in confronting a worldwide pandemic. 
}

\begin{multicols}{2}


\section*{Review of Meta-Analysis}

The infection fatality rate (IFR) of COVID-19 is one of the measures of disease impact that can be of importance for policy making. Over an extended period of time the CDC estimation of 0.65\% IFR \cite{CDC planning scenarios} was based on an unpublished meta-analysis \cite{Meyerowitz-Katz2020}. We have previously written a criticism of one of the 24 reports used by the meta-analysis \cite{Meyerowitz-Katz2020} based on the Diamond Princess cruise \cite{Russell2020,DPcritics}. Starting from an incorrectly estimated 1.3\% IFR for the Diamond Princess— the actual IFR for the Diamond Princess being 2.0\%—the report attempted to model the IFR in China and reported a value 0.6\%; however, the multiple errors in their calculations, including dividing by the same factor twice, render their results meaningless.

Here we discuss flaws in five additional studies cited in the meta-analysis. We note that these studies were not selected because of the presence of flaws, but instead they were sequentially chosen in each of the three categories in the meta-analysis and therefore can be taken as indicative of the quality to be expected. They include the following members of the three labeled categories in the meta-analysis: model estimates (first, on repatriation flights from Japan \cite{Nishiura2020}, second, on exported cases from China \cite{Jung2020}, third, on a study in France \cite{Salje2020}), observational studies (first, on cases in Beijing \cite{Tian2020}, second, on the Diamond Princess \cite{Russell2020} previously reviewed \cite{DPcritics}), and serological surveys (first, on a study of Santa Clara County \cite{Bendavid2020}). We discuss two types of problems: flaws in the studies per se, and flaws in the way in which they are used by the meta-analysis.

Characteristic problems in the cited studies include:
\begin{itemize}
    \item Small sample sizes
    \item Non-representative sampling (systematic biases)
    \item Incorrect definitions of symptomatic and asymptomatic  cases (identified and non-identified cases), typically assuming that cases which are non-symptomatic at the time of testing are the same as completely asymptomatic (never symptomatic) cases.
    \item Mathematical errors
    \item Errors of presentation that distort the meaning of findings
    \item No attempt to account for unattributed deaths from COVID-19 (e.g. if the infected individual died before being tested for COVID-19)
\end{itemize}

Characteristic problems in the meta-analysis include:
\begin{itemize}
    \item Misrepresentation of the IFR values in the original studies
    \item Inappropriate duplicate use of studies, or the information from studies, so that results averaged are not independent from each other, which gives uneven weight to the results of the original work.  
\end{itemize}

\section{}
{\em The Rate of Underascertainment of Novel Coronavirus (2019-nCoV) Infection: Estimation Using Japanese Passengers Data on Evacuation Flights}~\cite{Nishiura2020}.\\

At the end of January, Japan conducted three repatriation flights of Japanese nationals and their families on Jan 29-31 from Wuhan, China. Upon disembarkation from the planes, tests were performed of the passengers and the test results were used to estimate IFR in Wuhan. The study infers an IFR of  $0.3\% - 0.6\%$, based on a $3-6\%$ case fatality rate (CFR) for Wuhan and an estimated $\approx 9.2\%$ ratio of confirmed cases to infections. This estimate is obtained by taking the number of individuals who tested positive upon disembarkation, $8$, as a fraction of the total number of passengers, $565$, to be equal to the fraction of infected people detectable at the same time in the city of Wuhan. Given this estimate of the infection rate, the IFR is determined.

Flaws: (a) Small sample size. (b) Non-representative sample. (c) Nonsensical mathematical analysis. (d) No external validity to the ascertainment rate (percentage of infections that are tested and confirmed). (e) Ignoring confidence interval uncertainty.

Flaw (a): This study uses repatriation flights of Japanese nationals to estimate the prevalence in Wuhan. With $n=565$ passengers total, $8$ of them received positive test results. Thus, a sample size of only $8$ positive cases is used to infer the ascertainment rate, prevalence rate, and IFR in a population of 11 million. Given the highly heterogeneous nature of pandemic outbreaks in populations, a small sample size undermines conclusions. Moreover, the repatriation consisted of families rather than independent individuals, and the high correlation between cases among family members reduces the statistical independence of the sample.

Flaw (b): The use of repatriated flights of Japanese nationals to estimate the prevalence in Wuhan require two unjustified assumptions: (a) the $565$ passengers are a representative sample of the Wuhan population, and (b) the infections in Wuhan had no geographical, socio-economic, cultural, or other heterogeneity. The flight consisted of Japanese nationals and their families, indicating a distinct and potentially partially segregated population, with behavioral cultural differences relevant to disease transmission, compared to Chinese nationals generally. The general population and those who both fly on international flights and have Japanese citizenship should not be assumed to have the same prevalence due to distinct socioeconomic classes and social networks. We also note that geographical heterogeneity is a feature of this outbreak, and Japanese nationals in Wuhan cannot be assumed to be uniformly distributed geographically. The assumption by the paper of a homogeneous urban population is highly implausible.   

Flaw (c): The authors write an equation in which they set the cumulative number of infections in Wuhan (the confirmed cases divided by the ascertainment rate $c(t)/q$) divided by the total population of Wuhan $n$  times the detection window $T$ equal to the fraction of evacuees who tested positive ($8/565$)---i.e. 
\begin{equation}
\frac{8}{565}=\frac{c(t)T}{qn}.
\end{equation}
We immediately see a problem, however, which is that the left-hand side of the equation is dimensionless, while the right-hand side of the equation has units of time. This renders the equation meaningless. If, in fact, the sample were at all representative, an appropriate equation would be $\frac{8}{565}=\frac{c(t)}{qn}p$ where $p$ is the fraction of all infections in Wuhan that would be detectable at the time that the evacuees were tested. We will not carry through such an analysis here, however, since the entire concept of trying to estimate incidence in Wuhan from 8 positive cases among evacuees is fatally flawed.        

Flaw (d): The ascertainment rate that the authors derive is not valid, both due to a lack of a representative and independent sample and due to a mathematical error in their equation that renders their results meaningless. Even if it were correctly derived, however, it has no validity for application to other contexts, as it applies only to a particular snapshot in time when tests in China were just starting to ramp up. Any application of this rate to convert CFR to IFR under other conditions is therefore inappropriate and misleading (see Section II).

Flaw (e): The ascertainment rate, derived from 8 cases out of 565, comes with a 95\% CI range of $3.5-15.7$ positive results. The authors are evidently aware of this error range as they noted $q$ with $(5.0-20\%)$ 95\% CI, an error range that can half or double the point estimate of 10\%. Yet when calculating IFR subsequently, where IFR $=$ CFR $\times q$, they use 10\% for $q$ without addressing the error range to obtain their final estimation of China IFR of 0.3-0.6\% (China CFR estimated as 3-6\% at the time). Correctly including the error in both the China CFR estimate and the ascertainment rate would lead to a significantly larger error range for China’s IFR.  Of course, given all of the errors present in deriving the ascertainment rate, even a correctly calculated error range would be meaningless. 

In total, the multiple errors, including nonsensical mathematical equations, statistical uncertainty, non-transferability, and incorrect reporting of results, makes this paper unusable for estimating an IFR.

\section{}
{\em Real-Time Estimation of the Risk of Death from Novel Coronavirus (COVID-19) Infection: Inference Using Exported Cases}~\cite{Jung2020}.\\

This study modeled the epidemic growth rate in China. Its results are an estimate of CFR (more precisely the case fatality rate based upon cases and fatalities at one point in time during the outbreak, called cCFR) of $5-8\%$. The study did not report an IFR. The meta-analysis claimed an IFR of $0.5-0.8\%$ by multiplying the cCFR of the study by the ascertainment rate from the previously described Japanese passenger study in Section I.

Flaws: (a) Inappropriate duplicate use of studies. (b) Use of the ascertainment rate from the previous study is subject to all the flaws of that study (see Section I).

Flaw (a): There is no independent assessment of $q$ \cite{Jung2020}. We emphasize that this is a scientifically incorrect method of performing a meta-analysis. Since the ascertainment rate is a key determining factor for IFR, this is not an independent evaluation of the IFR as would be required for correct inclusion in a meta-analysis. Even if the ascertainment rate were valid, including multiple studies using the same ascertainment rate without independent validation of the value, as the meta-analysis did with both \cite{Nishiura2020} and \cite{Jung2020}, inflates the weight given to the data and assumptions from which the ascertainment rate in question was calculated.

Flaw (b): We note again that the ascertainment rate that is being used has multiple invalidating flaws detailed previously.

In total, this study has no independent assessment of the ascertainment rate which is fundamental for IFR estimation. The inclusion of both studies in the meta-analysis is methodologically flawed, and the fundamentally flawed ascertainment rate makes this study unusable for estimating an IFR.

\section{}
{\em Estimating the burden of SARS-CoV-2 in France} \cite{Salje2020}.\\

By assuming that the French and Diamond Princess cruise ship populations differ only in their age/sex composition and by assuming a particular distribution of infected individuals across age/sex groups, the authors estimate the IFR for each age/sex group in France. They report an overall IFR of 0.5\% (95\% credible interval: 0.3 to 0.9\%).

Flaws: (a) Misstatement of the study’s conclusions as being the population death rate when they only included hospital deaths. (b) Unjustified assumptions regarding contact rates and inaccurate analysis of the relationship between contact rates and probability of being infected. (c) Non-representative sample (Diamond Princess data) used to set the overall IFR. (d) No analysis of the sensitivity of the results to the assumed Bayesian prior distributions. (e) Meta-analysis flaw: inappropriate duplicate use of studies in using the Diamond Princess data. 

Flaw (a): In the abstract the authors state that “$0.5\%$ of those infected die” but later in their manuscript they note that all COVID-19 deaths occurring outside of hospitals—and in particular all retirement/nursing home populations—are excluded from their analyses. While it is valid to conduct an analysis that applies only to part of a population, it is misleading to report results derived from such analyses as if they apply to the population as a whole as the authors do in the abstract, particularly given that they exclude $37\%$ of the total COVID-19 deaths. (By May 7, the final date for data used in this study, there were $16,386$ deaths in hospitals in France and $9,604$ deaths that occurred outside of hospitals, including in retirement homes.~\cite{Salje2020}.)  

Flaw (b): While they have data on the number of deaths and hospitalizations in each age/sex group in France, they do not know how many people in each group were infected. By estimating the relative rates of infections among the age/sex groups, they are able to estimate the ratio of the IFRs between the groups. They assume that infections for each age/sex group are proportional to contact rate based on the supposed linearity of their Fig. S17. However, Fig. S17 actually has substantial curvature to it, bringing this assumption into question. Furthermore, the data they use to ascertain the contact rates of each of the age/sex groups is from a study conducted in 2012 rather than during the pandemic period. The authors then make a series of assumptions that are not validated by empirical data as to the effect of the lockdown on these contact rates in each age group. Different assumptions about the relative probability of infection in each age group will yield different results. For instance, when they perform a sensitivity analysis in which everyone from all age/sex groups is equally likely to be infected, their overall IFR is $0.7\%$ rather than $0.5\%$. 

Flaw (c): In order to estimate absolute IFRs for each age/sex group (rather than only the ratios between them), they also included in their Bayesian analysis the Diamond Princess cruise ship data, in which the number of infected people by age and sex is known. Thus, the absolute IFRs by age/sex group that they calculate are essentially derived from those on the Diamond Princess, using the French data to understand the relative effect of age/sex on IFR. The reason why their estimated overall IFR ($0.5\%$) is far lower than the IFR of $2\%$ on the Diamond Princess cruise ship is that their estimated age distribution of infected individuals in France is far younger than the age distribution of infected individuals on the Diamond Princess.

Their analysis is accurate only to the extent that the IFR for each age/sex group on the Diamond Princess equals the IFR for that age/sex group in France. This assumption is not justified; individuals on the Diamond Princess could substantially differ from individuals of the same age and sex in France. In particular, cruise ship passengers are, for their age, likely to be healthier than the general population. They also were likely to receive more rapid medical attention due to the active screening of the virus among the cruise ship passengers. Given that relatively small changes in age can substantially impact IFR (e.g. an age difference of ten years impacts the estimated IFRs by a factor of $3$ or $4$ according to this study), it seems likely that IFRs are also substantially impacted by relatively small differences in health or other variables. Recognizing that there may be a health difference between the populations, they perform a sensitivity analysis. However, their sensitivity analysis in which the Diamond Princess population has a $25\%$ lower IFR for each age/sex group is inadequate; it would not be surprising if the Diamond Princess IFRs for each particular age/sex group differed from their French counterparts by a greater factor. (The Diamond Princess data is also noisy, with only $15$ deaths total, although presumably this noise is accounted for in the wide credible intervals that the authors report.)  

Flaw (d): Finally, their paper is based on Bayesian analysis, meaning that the authors had to select a prior distribution for each of their parameters. However, they provide no justification for their choice of priors, nor is any sensitivity analysis for the priors reported. The robustness/usefulness of their estimates is thus left unjustified, and it is unclear how their results would differ had they chosen a different set of priors.

We now consider how an IFR estimate could be adjusted to include the non-hospitalized retirement home deaths.  There were $16,384$ total hospitalized deaths and $9,604$ non-hospitalized deaths.  We have that 
$$\text{IFR}_{\text{total}} = 
\frac{\text{Total Deaths}} {\text{Total Infections}}= $$
$$\frac{16384+9604}{16384/\text{IFR}_{\text{Hospitalized}}+\text{NHI}}=
\frac{1.59 \text{~IFR}_{\text{Hospitalized}}}{1+\frac{\text{NHI}}{16384/\text{IFR}_{\text{Hospitalized}}}}$$  where NHI is the number of non-hospitalized infections.  Assuming the number of non-hospitalized infections is small compared to the total number of infections in France, an IFR that includes deaths in retirement homes can be calculated from an IFR that does not by multiplying by 1.59. (This factor is so large in part because individuals in nursing homes were more likely to be infected.) Thus, an estimate of $0.5\%$ becomes $0.8\%$, and an estimate of $0.7\%$ becomes $1.1\%$. The study’s $95\%$ credible interval for the IFR of $0.3$ to $0.9\%$ becomes $0.6$ to $1.6\%$ when non-hospitalized deaths are included. However, this credible interval does not include the additional sources of error discussed above, including the systematic bias that a healthier Diamond Princess population would lead to IFR estimates that are too low, which render it impossible to obtain an accurate estimate of IFR from this study.

Flaw (e): Meta-analysis flaw: By including both the Diamond Princess study\cite{Russell2020} and this study that estimated IFR based on the Diamond Princess data, the meta-analysis effectively includes IFR estimation of the Diamond Princess data twice, a similar meta-analysis error as found in Section II.

\section{}
{\em Characteristics of COVID-19 infection in Beijing}~\cite{Tian2020}.\\

This is an observational study of patients from Beijing, with $262$ confirmed cases that were transferred to designated hospitals and $3$ deaths. It is used as $1.15\%$ IFR in the meta-analysis.

Flaws: (a) non-representative population sample, (b) Meta-analysis flaw: misstatement of the study results by the meta-analysis.

Flaw (a): The study cannot be considered to contain a representative sample of the population. It over-represents travelers (from Wuhan) consisting of $26\%$ of the sample. Without widespread transmission in the population, those infected cannot be considered to be representative of the population as a whole. In particular, intra-city travelers are young and tend to have reduced chances of prior health conditions. Due to homophily, those that they infect in the first generation of infections would also not be representative. 

Flaw (b): The abstract of the Beijing study reads ``the fatality of COVID-19 infection in Beijing was $0.9\%$" \cite{Tian2020}. The $0.9\%$ was obtained by taking $3$ deaths divided by $342$ total confirmed cases reported in Beijing by Feb 10, 2020 \cite{Tian2020}, including $80$ cases that were not transferred to designated hospitals. Thus the ``0.9\% fatality rate'' is the CFR, not IFR. Thus, the meta-analysis \cite{Meyerowitz-Katz2020} misstated the CFR as IFR. Moreover the meta-analysis divides the $3$ deaths by the $262$ confirmed cases that were transferred to designated hospitals to get a CFR of $1.15\%$, which they include as the IFR.

\section{}
{\em COVID-19 Antibody Seroprevalence in Santa Clara County, California}~\cite{Bendavid2020}.\\

This study was one of the first seroprevalance studies in the US, which estimated a $2.8\%$ prevalence rate, meaning infections were more than $50$ times higher than the confirmed cases in Santa Clara county at the time of the study. The estimated IFR was $0.17\%$. This is one of the lowest IFR estimations the meta-analysis cited.

Flaws: (a) Non representative sample. (b) Widely criticized sensitivity and specificity of the tests.

Flaw (a): The population sample was obtained by on-line sign-up without measures to determine whether or not it was a representative sample. This is a highly suspect sampling method that should not be used in a meta-analysis. 

Flaw (b): Sensitivity and specificity: The serological test was not approved by the FDA and was widely criticized \cite{SCcritics1, SCcritics2}. 

Criticisms of this study have been present since the day it was published in pre-print form. The sampling method, the statistical tests, the sensitivity/specificity of the tests, are all reported to be seriously flawed \cite{SCcritics1, SCcritics2}. The population prevalence estimate, the ratio of IFR to CFR used to determine the IFR, which is the basis of this study is wildly inconsistent with other studies. Considering these criticisms, the inclusion of it in a meta-analysis is inappropriate. As one of the lowest IFRs in the meta-analysis, it had an outsized influence over the reported result.

We note that one of the authors of this Santa Clara study has also posted a meta-analysis (which includes this Santa Clara study)~\cite{M3}. The author has been faulted for not reporting conflict of interest due to funding from a financially interested party and for advocating for policies consistent with that party’s view~\cite{C1}. The severity of misconduct and its damaging effects are described in the report.  

\section*{Summary}

Our review of the meta-analysis points out multiple flaws in the papers cited and in their use by the meta-analysis. In addition to these specific flaws, none of the studies discussed above attempted to estimate the number of deaths from COVID-19 that were not attributed to COVID-19, leading to systematic underestimates of the IFR, as noted in the meta-analysis. Excess mortality data \cite{Rossen2020} suggests a significant number of unattributed deaths from COVID-19, although some of the excess mortality may be from indirect effects of the pandemic. (Such deaths should still be considered when evaluating the overall impact of the pandemic.) 

It is also important to note that the IFR is not a biological invariant but rather significantly varies with the distribution of age, health and other qualities of those infected as well as what type of medical care they receive.  Thus, the IFR calculated for a particular population of infected individuals is not immediately generalizable to other populations.  Meta-analyses that consider IFR as a well-defined quantity and that average values from various populations to produce a single estimate without accounting for such differences are therefore problematic.  (The lead author has recently posted an additional meta-analysis that investigates age-stratified IFRs \cite{Levin2020}; such an analysis is an improvement but still does not take into account other factors such as differences in the underlying health of each age group and differences in access to treatment.)

In summary, we have pointed out flaws in 5 papers included in the meta-analysis \cite{Meyerowitz-Katz2020}, as well as the methods of their use in the meta-analysis itself. We previously published criticism of one of the other reports which the meta-analysis was based on \cite{Russell2020, DPcritics}. Inclusive, of 6 studies that we analyzed, not one provided a sound statistical and methodological basis for an IFR determination. With all of its flaws, we are surprised to see that this meta-analysis has been used in preprint form as the CDC’s estimate for the IFR of SARS-CoV-2 \cite{CDC planning scenarios} and that it is provided any credibility by the scientific community.

\end{multicols}


\begin{thebibliography}{20}

\bibitem{CDC planning scenarios} CDC. COVID-19 Pandemic Planning Scenarios. \href{https://www.cdc.gov/coronavirus/2019-ncov/hcp/planning-scenarios.html}{https://www.cdc.gov/coronavirus/2019-ncov/hcp/planning-scenarios.html} (accessed November 18, 2020).

\bibitem{Meyerowitz-Katz2020} Meyerowitz-Katz G, Merone L. A systematic review and meta-analysis of published research data on COVID-19 infection-fatality rates. medRxiv 2020; \href{https://doi.org/10.1101/2020.05.03.20089854}{DOI: 2020.05.03.20089854}.

\bibitem{Russell2020} Russell T, Hellewell J, Jarvis C, et al. Estimating the infection and case fatality ratio for COVID-19 using age-adjusted data from the outbreak on the Diamond Princess cruise ship. medRxiv 2020; \href{https://doi.org/10.1101/2020.03.05.20031773}{DOI: 10.1101/2020.03.05.20031773}.

\bibitem{DPcritics} Shen C, Van Gennep D, Bar-Yam Y. The IFR of the Diamond Princess has been misreported, best current value is 2.0\%. New England Complex Systems Institute \url{https://necsi.edu/the-ifr-of-the-diamond-princess-has-been-misreported-best-current-value-is-2} (accessed November 18, 2020).

\bibitem{Nishiura2020} Nishiura H, Kobayashi T, Yang Y, et al. The Rate of Underascertainment of Novel Coronavirus (2019-nCoV) Infection: Estimation Using Japanese Passengers Data on Evacuation Flights. Journal of Clinical Medicine 2020; 9(2), 419. \href{https://doi.org/10.3390/jcm9020419}{DOI: 10.3390/jcm9020419}.

\bibitem{Jung2020} Jung S, Akhmetzhanov AR, Hayashi K, et al. Real-Time Estimation of the Risk of Death from Novel Coronavirus (COVID-19) Infection: Inference Using Exported Cases. Journal of Clinical Medicine 2020; 9(2), 523. \href{https://doi.org/10.3390/jcm9020523}{DOI: 10.3390/jcm9020523}.

\bibitem{Salje2020} Salje H, Tran Kiem C, Lefrancq N, et al. Estimating the burden of SARS-CoV-2 in France. Science 2020; 369(6500), eabc3517. \href{https://doi.org/10.1126/science.abc3517}{DOI: 10.1126/science.abc3517}

\bibitem{Tian2020} Tian S, Hu N, Lou J, et al. Characteristics of COVID-19 infection in Beijing. Journal of Infection 2020; 80(4), 401–406. \href{https://doi.org/10.1016/j.jinf.2020.02.018}{DOI: 10.1016/j.jinf.2020.02.018}.

\bibitem{Bendavid2020} Bendavid E, Mulaney B, Sood N, et al. COVID-19 Antibody Seroprevalence in Santa Clara County, California. medRxiv 2020; \href{https://doi.org/10.1101/2020.04.14.20062463}{DOI:10.1101/2020.04.14.20062463}.

\bibitem{SCcritics1} Kreiger L. Stanford coronavirus research: Did politically-motivated scientists hype their speedy study? The Mercury News, May 24, 2020.  \url{https://www.mercurynews.com/2020/05/24/coronavirus-research-stanford-scientists-accused-of-hyping-covid-19-antibody-study/} (accessed November 18, 2020).

\bibitem{SCcritics2} McCormick E. Why experts are questioning two hyped antibody studies in coronavirus hotspots. The Guardian, April 23, 2020.  \url{https://www.theguardian.com/world/2020/apr/23/coronavirus-antibody-studies-california-stanford} (accessed November 18, 2020).

\bibitem{M3} Ioannidis JPA. Infection fatality rate of COVID-19 inferred from seroprevalence data. Bulletin of the World Health Organization 37, 2020.  \url{https://www.who.int/bulletin/online_first/BLT.20.265892.pdf} (accessed November 18, 2020).

\bibitem{C1} Lee SM. JetBlue's Founder Helped Fund A Stanford Study That Said The Coronavirus Wasn't That Deadly. Buzzfeed May 15, 2020. \url{https://www.buzzfeednews.com/article/stephaniemlee/stanford-coronavirus-neeleman-ioannidis-whistleblower} (accessed November 18, 2020).

\bibitem{Rossen2020}Rossen LM, Branum AM, Ahmad FB, Sutton P, Anderson RN. (2020). Excess Deaths Associated with COVID-19, by Age and Race and Ethnicity — United States, January 26–October 3, 2020. MMWR Morb Mortal Wkly Rep 2020; 69: 1522–7. \url{https://www.cdc.gov/mmwr/volumes/69/wr/mm6942e2.htm} (accessed November 18, 2020).

\bibitem{Levin2020} Levin AT, Hanage WP, Owusu-Boaitey N, Cochran KB, Walsh SP, Meyerowitz-Katz G. Assessing the Age Specificity of Infection Fatality Rates for COVID-19: Systematic Review, Meta-Analysis, and Public Policy Implications. medRxiv 2020; \href{https://doi.org/10.1101/2020.07.23.20160895}{DOI: 10.1101/2020.07.23.20160895}












\end{thebibliography}

\end{document}